# Using Car-Parrinello simulations and microscopic order descriptors to reveal two locally favored structures with distinct molecular dipole moments and dynamics in ambient liquid water.

Ioannis Skarmoutsos [a,*], Giancarlo Franzese [b,*] and Elvira Guardia [c,*]

[a] Laboratory of Physical Chemistry, Department of Chemistry, University of Ioannina, 45110 Ioannina, Greece

[b] Secció de Física Estadistica i Interdisciplinària, Departament de Física de la Matèria Condensada & Institut de Nanociència i Nanotecnologia (IN2UB), Universitat de Barcelona, C. Martí i Franquès 1, 08028 Barcelona (Spain)

[c] Departament de Física, Universitat Politècnica de Catalunya, Campus Nord-Edifici B4-B5, Jordi Girona 1-3, Barcelona E-08034, Spain.

Water is essential for life and technological applications, mainly for its unique thermodynamic and dynamic properties, often anomalous or counterintuitive. These anomalies result from the hydrogen-bonds fluctuations, as evidenced by studies for supercooled water. However, it is difficult to characterize these fluctuations under ambient conditions. Here, we fill this knowledge gap thanks to the Car-Parrinello *ab initio* molecular dynamics (MD) simulation technique. We calculate the local structural order parameter $\zeta$, quantifying the coordination shells separation, and find two locally-favored structures or states: High-$\zeta$ and Low-$\zeta$. On average, High-$\zeta$ molecules have a tetrahedral arrangement, with four hydrogen bonds, and the first and the second coordination shell well separated. The Low-$\zeta$ molecules are less connected, partially merging the first and the second shells. The appearance of isosbestic points in the radial distribution functions and the collective density fluctuations at different length scales and timescales reveal that the two-state model, consistent with available experimental data for supercooled water, also holds under ambient conditions, as we confirm by analyzing the vibrational spectrum of both types of water molecules. Significant consequences of the structural differences between the two states are that High-$\zeta$ molecules have a dipole moment 6 % higher than Low-$\zeta$. At the same time, Low-$\zeta$ structures are more disordered and with more significant angular fluctuations. These differences are also reflected in the dynamics under ambient conditions. The Low-$\zeta$ molecules decorrelate their reorientation faster than High-$\zeta$ and merge their coordination shells within 0.2 ps, while the High-$\zeta$ preserve the shell separation for longer times. Our analysis shows that first-principle calculations make predictions under ambient conditions calling for new and definitive experiments confirming the two-state model.

[*] Corresponding author emails: iskarmoutsos@hotmail.com, gfranzese@ub.edu, elvira.guardia@upc.edu

1. Introduction

The complex behavior of water has motivated the scientific community to develop novel theories and methodologies to provide deeper insight into the molecular causes of its anomalies compared to most liquids[1-11]. A molecular understanding of the water properties is crucial in numerous processes in chemistry, physics, geosciences, biology, and life evolution[3,4,9,10, 12-17].

Frequently, the water anomalies are associated with the heterogeneous structure and dynamics of the liquid[18-22], although this view is strongly debated[23-27]. Several translational and orientational order parameters help in characterizing the heterogeneity[28-39]. These approaches seek locally favored arrangements with low configurational entropy and investigate their impact on the thermodynamics and dynamics of water.

Appropriate microscopic structural descriptors for the local translational and orientational order can reveal critical structural features in liquids[40]. For example, Russo and Tanaka introduced a structural parameter accounting for the intermolecular hydrogen bonding and describing the degree of translational order in the second coordination shell[36,37,41-43]. They identified two distinct local structures, with or without a well-formed second coordination shell. When the first shell has hydrogen bonds with high tetrahedral order, the descriptor calculated on the second shell has a high value. On the other hand, if the first and the second shell partially merge, the tetrahedral order around the central molecule decreases, and the descriptor has a low value.

Also, previous *ab initio* molecular dynamics (MD) simulations, based on density functional theory (DFT), revealed that water molecules under ambient conditions could be classified into two different types, depending on the tetrahedral order of their local environment[35,44]. The observation of structural fluctuations in ambient liquid water via DFT-based MD simulations is consistent with recent experiments[45] on water ultrafast vibrational motion, pointing out the importance of the quantum mechanical nature of water hydrogen bonds at ambient conditions. Here, our main aim is to study the dynamics of the water molecules in the liquid state, calculating the local structural descriptor proposed by Russo and Tanaka with quantum mechanical methodologies to bring more profound insight into water's peculiar properties.

## 2. Computational Methods

We calculate the properties of liquid water by analyzing the trajectory of a Car-Parrinello MD (CPMD) simulation of $N$=96 water molecules in a cubic box with a fixed density (0.997 g/cm$^3$) and periodic boundary conditions. We use the BLYP density functional[46,47], with a plane-wave cutoff set at 80 Ry. Also, we employ dispersion-corrected atom-centered pseudopotentials (DCACPs)[48,49] in the Troullier-Martins format[50] for oxygen and hydrogen to effectively treat the dispersion interactions. Previous studies[49] have shown that using these pseudopotentials successfully reproduces water's structural and dynamic properties and the reference interaction energies obtained by high-level *ab initio* calculations. We use a fictitious electronic mass parameter of 400 a.m.u. to study the time evolution of the electronic degrees of freedom. The system is pre-equilibrated for 200 ps by performing a classical MD simulation, using a flexible water model parametrized using *ab initio* data[51]. Next, following the previous studies[49], we equilibrate for 3 ps with CPMD at constant $N$, volume $V$, and temperature $T$=330 K. Setting 330 K avoids falling into the temperature range where CPMD simulations with the BLYP functional suffer from non-ergodic behavior on timescales shorter than 20 ps[52]. Next, we calculate the observables for 15 ps with CPMD in the microcanonical ensemble (constant $N$, $V$, and energy $E$). Previous systematic studies[44] about the effects of finite size and simulation timescale (see the Supporting Information of Ref. 44 for more details) show that this specific simulation protocol accurately describes the water local structure and dynamics over the picoseconds' timescale.

To investigate the local structural order in liquid water, we calculate the descriptor $\zeta$ introduced by Russo and Tanaka[36,37,41-43] and its coarse-grained[41] average, $\zeta_{CG}$, over the first coordination shell of each molecule. The $\zeta$ of a water molecule is, by definition, the difference between its distance from the closest non-hydrogen-bonded neighbor and the furthest hydrogen-bonded neighbor. We give examples of configurations where this order parameter is positive or negative in Section 3.2. In the case of ambient or supercooled liquid water, large positive values of $\zeta$ indicate a sharp separation between the first two coordination shells. On the other hand, negative or vanishing values of $\zeta$ correspond to the second coordination shell merging with the first.

In the case of fluid water at very elevated pressures and supercritical temperatures, where the first coordination shell is very densely packed and consists of 7-14 water molecules[53], negative values of $\zeta$ indicate an interpenetration of the short-range local HB network of the individual water molecules by non-hydrogen-bonded neighbors. In such cases, the change in the sign of the descriptor $\zeta$ from positive to negative values has successfully provided accurate identifications of the structural transitions associated with the Frenkel line[53], further verifying its suitability in describing several different structural phenomena taking place in water in a vast range of thermodynamic conditions.

To identify hydrogen-bonded water pairs, we employ the geometric criterion used in previous CPMD studies[44], which has been successfully tested for ambient liquid water. According to this criterion, a hydrogen bond between two water molecules exists if the intermolecular distances are $R_{O...O} \leq 3.38$ Å, $R_{H...O} \leq 2.48$ Å, and the donor-acceptor angle quantifying the deviation of the hydrogen bond from linear is $\varphi \equiv H - \hat{O}...O \leq 30°$ (where the notations '…' and '–' denote intermolecular and intramolecular vector, respectively). A previous study by Shi and Tanaka[37] shows that the calculation of $\zeta$ is robust for several standard hydrogen bonding criteria.

**3. Results and Discussion**

**3.1 Two-State Local Intermolecular Structure**

As shown in the previous works[36,47,41-43], one can decompose the probability density distributions for $\zeta$ and $\zeta_{CG}$ into two Gaussians (Figure 1). We test that a single-Gaussian fit has a reduced-$\chi^2$ statistic 18 times larger than the two-Gaussians fit and a relative coefficient of determination, $R^2$, 0.17% smaller. These differences are more significant at lower temperatures[36,37,41-43]. For $\zeta$, the first Gaussian has a maximum at $\zeta \approx$ -0.02 Å, a vanishing value consistent with a similar analysis for TIP4P/2005, TIP5P, and ST2 water[43], corresponding to a considerable shell interpenetration. The second Gaussian has a peak at $\zeta \approx 0.37$ Å, characteristic of a sharp shell separation. Note also that the intersection of the two Gaussians is located at $\zeta \approx 0.0$ Å. Hence, ambient water displays two populations with distinct local structures, low-$\zeta$ and high-$\zeta$.

For $\zeta_{CG}$, the two Gaussian maxima are located at $\zeta_{CG} \approx 0.07$ Å and 0.46 Å, i.e., 0.1 Å higher than those for $\zeta$. Hence, also the coarse-grained average displays two components. However, the Gaussians for $\zeta_{CG}$ have a more extensive overlap than the $\zeta$ case. To characterize the $\zeta_{CG}$ components, we plot parametrically in Figure 1 the two quantities $\zeta_{CG}$ and $\zeta$, calculated for each water molecule, and observe that they are linearly related with a coefficient of $\approx 1$ and a constant of $\approx 0.1$ ($\zeta_{CG} = 0.98\ \zeta + 0.14$). Hence, molecules with $\zeta_{CG}$ up to its first Gaussian maximum (at $\approx 0.1$ Å), mostly belonging to the low-$\zeta_{CG}$ component, have $\zeta \leq 0$, i.e., they also belong to the low-$\zeta$ population. Therefore, a low-$\zeta$ molecule, i.e., having its first and second shells merging, also has first neighbors with a similar disordered structure. Consequently, the structural inhomogeneity around a molecule has a spatial correlation extending over at least two coordination shells.

The low-$\zeta$ Gaussian has a tail that extends approximately up to the maximum of the high-$\zeta$ Gaussian at $\zeta \approx 0.37$ Å. Therefore, we set $\zeta_H = 0.37$ Å as the threshold above which molecules are, by definition, *High-$\zeta$*, while the rest are *Low-$\zeta$*. The linear relation between $\zeta_{CG}$ and $\zeta$ shows that High-$\zeta$ molecules have $\zeta_{CG} \geq 0.5$Å. Our classification gives that 36% of the molecules are High-$\zeta$ and the rest 64% are Low-$\zeta$, respectively.

To provide a quantitative description of their local structures, we calculate the oxygen-oxygen (O-O) intermolecular pair radial distribution function (RDF, $g^{OO}(r)$) and the coordination number $N_C^{OO}$ when the central molecule is High-$\zeta$ or Low-$\zeta$ (Figure 2). From our previous works[44], we know that the average RDFs for O-O, O-H, and H-H, calculated by CPMD simulations, are reliable and in agreement with available experimental data[54].

The High-$\zeta$ $g^{OO}(r)$ exhibits a well-formed, deep first minimum at $r \approx 3.23$ Å, marking a sharp separation between the first and the second coordination shell. On the other hand, the Low-$\zeta$ $g^{OO}(r)$ has a smoother shape with a shallow minimum at $r \approx 3.48$ Å, followed by a broad second maximum, consistent with partially merging shells.

By integrating the RDF up to the radial distance $r$, we calculate $N_C^{OO}(r)$. For High-$\zeta$ molecules, $N_C^{OO}(r)$ flattens out at r ≈ 3.23 Å, reaching a value of $N_C^{OO}(r)$ ≈ 3.85, as expected for a tetrahedral shell. For Low-$\zeta$ molecules, $N_C^{OO}(r)$ within the first RDF peak, at $r$ ≈ 3.48 Å, grows smoothly up to $N_C^{OO}(r)$ ≈ 5.36, consistent with two partially merged shells with a non-tetrahedral arrangement.

The functions in Figure 2 are evaluated considering only the state of the central molecule. We get further information on the local structural heterogeneities in ambient liquid water by calculating the quantities in which we select the state of both molecules in the pair. Hence, we evaluate the O-O RDFs for High-$\zeta$-High-$\zeta$, High-$\zeta$-Low-$\zeta$, and Low-$\zeta$-Low-$\zeta$ pairs of molecules and the corresponding coordination numbers (Figure 3).

We find that the three RDFs are pretty different. The High-$\zeta$-High-$\zeta$ RDF has a first peak approximately twice as high as the other two RDFs, with a minimum at r ≈ 3.28 Å. Hence highly tetrahedral molecules are more likely to cluster in a regular structure than those with lower tetrahedral order. Furthermore, although with less probability, they form well separated first and second shells also when their neighbors are Low-$\zeta$ molecules, as shown by the High-$\zeta$-Low-$\zeta$ RDF, which has a deep minimum at r ≈ 3.23 Å separating two maxima that are less intense than those of the High-$\zeta$-High-$\zeta$ RDF. Nevertheless, while the High-$\zeta$ molecules have a structure that correlates at least over two coordination shells, the Low-$\zeta$ decorrelate at a much shorter distance, as demonstrated by the Low-$\zeta$-Low-$\zeta$ RDF. Indeed, as expected for merging coordination shells, the first RDF maximum for the Low-$\zeta$ molecules is broad and separated only by a shallow minimum, at r ≈ 3.48 Å, from a mild second maximum.

The integration of the three different RDFs, calculating the coordination numbers, gives us further information about the local structures around High-$\zeta$ and Low-$\zeta$ molecules. We find that at ambient conditions, a High-$\zeta$ molecule has in its first coordination shell, on average, less than two High-$\zeta$ and slightly more than two Low-$\zeta$ molecules (Figure 3). Instead, a Low-$\zeta$ molecule has approximately five molecules in its first coordination shell, of which about four are Low-$\zeta$, and one is High-$\zeta$ (Figure 3).

We quantitatively estimate the average composition of the first coordination shell of each type of molecule by calculating the local mole fractions[55]

$$X_A^{Shell-of-A}(r) = \frac{N_{AA}(r)}{N_{AA}(r) + N_{AB}(r)}, \quad X_A^{Shell-of-B}(r) = \frac{N_{BA}(r)}{N_{BA}(r) + N_{BB}(r)} \quad (1)$$

$$X_B^{Shell-of-A}(r) = \frac{N_{AB}(r)}{N_{AB}(r) + N_{AA}(r)}, \quad X_B^{Shell-of-B}(r) = \frac{N_{BB}(r)}{N_{BA}(r) + N_{BB}(r)}. \quad (2)$$

Here $A$ and $B$ stand for High-$\zeta$ and Low-$\zeta$, the first coordination shell distances are $r=3.23$ Å and 3.48 Å for High-$\zeta$ and Low-$\zeta$, respectively, and, for notation simplicity, we drop the $OO$ apex from the coordination numbers symbol. We find $X_{High}^{Shell-of-High}=0.45$, $X_{Low}^{Shell-of-High}=0.55$, $X_{High}^{Shell-of-Low}=0.27$, and $X_{Low}^{Shell-of-Low}=0.73$. Therefore, each type of molecule has a higher component of the same type in its first coordination shell than in bulk, where 36% and 64% of molecules are High-$\zeta$ and Low-$\zeta$, respectively, as found above.

Having a High-$\zeta$ molecule, on average, 3.85 first neighbors and a Low-$\zeta$ molecule 5.36, these mole fractions give that the High-$\zeta$ first coordination shell has 1.73 High-$\zeta$ neighbors and 2.12 Low-$\zeta$. On the other hand, the Low-$\zeta$ shell has 1.45 High-$\zeta$ and 3.91 Low-$\zeta$ first neighbors, consistent with the description above.

An intriguing feature of our RDFs is that they exhibit crossing points located at about 2.9, 3.8, and 5.15 Å, which coincide with the isosbestic points of the experimental RDFs reported in the literature for both liquid $D_2O$ and $H_2O$ by changing temperature or pressure[56]. Such an isosbestic point is usually related to the presence in the liquid of two components, or species, that contribute independently to the measured quanties[56,57]. In the case of water, the isosbestic point is associated with the existence of two states that differs in structure and other properties[56]. Different formulations of the two-state models have been proposed so far, showing that they can reproduce, at least qualitatively, all the anomalies of water and the different thermodynamic scenarios that have been introduced to rationalize them[2]. Therefore, our results show that water under ambient conditions can be well described by combining two states, the High-$\zeta$ and the Low-$\zeta$ components.

## 3.2 Hydrogen Bonding and Local Orientational Order

We estimate the fraction of High-$\zeta$ and the Low-$\zeta$ water molecules forming hydrogen bonds (Figure 4). We found that ~5% of High-$\zeta$ molecules form two hydrogen bonds, ~16% three, ~72% four, and ~6% five. On the other hand, ~2% of Low-$\zeta$ molecules form one hydrogen bond, ~14% two, ~39% three, ~41% four, and ~3% five. Hence, most High-$\zeta$ molecules participate in a tetrahedral hydrogen-bond network without defects, while the opposite is true for the Low-$\zeta$ molecules. Overall, the average number of hydrogen bonds is 3.8 for High-$\zeta$ and 3.3 for Low-$\zeta$ molecules. Interestingly, this average number for a High-$\zeta$ molecule equals its first-shell coordination number. Hence, High-$\zeta$ molecules are hydrogen-bonded to all the four neighbors in their first coordination shell under ambient conditions.

To provide deeper insight into the orientational ordering of the nearest neighbors around the High-$\zeta$ and the Low-$\zeta$ water molecules, we calculate their tetrahedral[32,33] ($q_{tetr}$) and trigonal[33] ($q_{trig}$) orientational order parameters. The tetrahedral order parameter[32] is defined as

$$q_{tetr} = 1 - \left\langle \frac{3}{8} \sum_{j=1}^{3} \sum_{k=j+1}^{4} \left( \cos\varphi_{jik} + \frac{1}{3} \right)^2 \right\rangle. \qquad (3)$$

Its value ranges between 0, for a disordered configuration, and 1, for a perfect tetrahedral arrangement. Here, $\phi_{jik}$ is the angle formed by the vectors $\vec{r}_{ij}$ and $\vec{r}_{ik}$ connecting the oxygen of the central molecule $i$ to those of two of the four nearest neighbors, $j$ and $k$, and the brackets stand for the thermodynamic average over the entire system. The trigonal order parameter[33] $q_{trig}$ is defined as

$$q_{trig} = 1 - \left\langle \frac{4}{7} \sum_{j=1}^{2} \sum_{k=j+1}^{3} \left( \cos\varphi_{jik} + \frac{1}{2} \right)^2 \right\rangle \qquad (4)$$

for three closest neighbors of a central molecule $i$. This parameter reaches its maximum, 1, for a perfect trigonal arrangement, and the minimum, 0, for the disordered case.

We find that the High-$\zeta$ molecules have a normalized probability distribution for $q_{tetr}$ that is strongly asymmetric with a maximum at $q_{tetr} \approx 0.86$ and an average of $\langle q_{tetr} \rangle \approx 0.75$, confirming that they have a well-developed tetrahedral order (Figure 5). The Low-$\zeta$ molecules have a broad distribution that flattens out from $q_{tetr} \approx 0.53$ to $0.80$ with an average of $\langle q_{tetr} \rangle \approx 0.64$, consistent with a more disordered arrangement.

On the other hand, we find that the trigonal order parameter is almost unable to discriminate between the two local structures (Figure 5). The differences in its normalized probability distribution between the High-$\zeta$ and the Low-$\zeta$ case are less pronounced, as also reflected by the average values of $\langle q_{trig} \rangle \approx 0.78$ and $0.70$ for the 'High-$\zeta$ and the Low-$\zeta$ molecules, respectively.

Water molecules in the High-$\zeta$ state participate in a symmetrical tetrahedral local HB network (Figure 6a). The first coordination shell is well-separated from the second, leading to a positive value of $\zeta$, directly related to the shells' separation distance. On the other hand, those in the Low-$\zeta$ state have the first two coordination shells partially merged, with a fifth neighbor in the first (Figure 6b), consistent with the O-O RDF (Figure 2a). The interpenetration of H-bonded and non-H-bonded neighbors causes a negative value of $\zeta$.

**3.3 Dipole Moment**

To further characterize the different local structures found in ambient liquid water, we consider the dipole moment, which is a quantity affected by the local environment. We estimate the molecular dipole moment by calculating the maximally localized Wannier functions[58,59]. We employ these functions to partition the total charge density of the system into single-molecule contributions and calculate the dipole moments of each molecule from the ionic and the Wannier function center positions[60-62]. Due to the use of pseudopotentials, we consider explicitly per each water molecule 8 electrons, 4 doubly occupied Wannier functions, for the spin degeneracy and their relative centers[62].

We find that the average dipole moment over all the water molecules in the liquid is 2.97 D, in excellent agreement with the experimental value of 2.90 D[63]. However, the High-$\zeta$ molecules exhibit a normalized probability distribution function for the molecular dipole moments centered at higher values than the Low-$\zeta$ (Figure 7). The average dipole moment for the High-$\zeta$ molecules is 3.08 D, about 6% higher than the Low-$\zeta$ value of 2.91 D, and very close to the value of 3.09 D reported for ice Ih with a strong tetrahedral symmetry[64].

Hence, our calculation of the dipole moments confirms that the High-$\zeta$ structure is tetrahedral, while the Low-$\zeta$ is more disordered.

Previous works revealed that the 2D:1A and 2A:1D dipole moments are almost equal when using the Wannier function center positions[65] and that, depending on the method used to localize the electron density around the water molecules, the dipole moments calculations can exhibit small variations[66]. However, as mentioned above, we find that the average dipole moment over all the water molecules in the liquid is in excellent agreement with the experimental value. Therefore, we consider our calculations for the water dipole moments reliable.

Previous studies[35,67] investigated the dependence of the dipole moment of the water molecules, at ambient conditions, on the number of HBs they form. Here we observe that the average dipole moment of the High-$\zeta$ and Low-$\zeta$ molecules are approximately equal to those of molecules forming four and three HBs (3.07 D and 2.89 D), respectively[35]. This observation is consistent with the result that the average number of HBs per High-$\zeta$ and Low-$\zeta$ water molecule is close to four (3.8) and three (3.3), respectively, as shown in Section 3.2.

**3.4 Reorientational Dynamics**

To characterize the reorientation dynamics of the water molecules in different environments, we calculate the time-dependent angular Van Hove self-correlation functions[67]

$$G(\theta,t) = \frac{2}{N \cdot \sin\theta} \sum_{i=1}^{N} \left\langle \delta\left[\theta - \cos^{-1}\left(\vec{u}_i(0) \cdot \vec{u}_i(t)\right)\right]\right\rangle, \tag{5}$$

where $\vec{u}_i$ is a unitary intramolecular vector associated with a molecule $i$ and $t$ is the time, for molecules that at $t=0$ are High-$\zeta$ or Low-$\zeta$. This quantity probes the angular self-fluctuations of the molecules over different timescales $t$. We average $1/2 \cdot \sin\theta \cdot G(\theta,t)$ over the two O-H intramolecular vectors for three characteristic times, $t=0.1$, 0.2, and 0.3 ps (Figure 8). We find that Low-$\zeta$ water molecules have angular fluctuations that are more populated at high values than those for High-$\zeta$ and decorrelate in time faster than the High-$\zeta$ molecules.

Also, for the same vectors $\vec{u}_i$ we calculate the Legendre reorientation time-correlation functions (TCF)

$$C_{\ell R}(t) = P_\ell \left\langle \vec{u}_i(0) \cdot \vec{u}_i(t) \right\rangle \qquad (6)$$

where $P_\ell$ is a Legendre polynomial of order $\ell$, for molecules that at $t=0$ are High-$\zeta$ or Low-$\zeta$. We consider the first, the second, and the third-order Legendre polynomials and average over the two O–H vectors (Figure 9).

We estimate the first, second, and third-order Legendre orientation correlation times, $\tau_{1R}$, $\tau_{2R}$, and $\tau_{3R}$, for water molecules that at $t=0$ are High-$\zeta$ or Low-$\zeta$ (Table 1). For each $C_{\ell R}(t)$ we calculate

$$\tau_{\ell R} = \int_0^\infty C_{\ell R}(t) \cdot dt \qquad , \qquad \ell = 1-3 \qquad (7)$$

after fitting the correlation function with a double exponential for simplicity. We find that are High-$\zeta$ molecules are slower than Low-$\zeta$. Therefore, the structural difference of the first coordination shell leads to different orientational dynamics under ambient conditions. The more disordered Low-$\zeta$ water molecules decorrelate faster than the tetrahedrally-ordered High-$\zeta$ molecules.

Also, we compare the ratios $\tau_{1R}/\tau_{2R}$ and $\tau_{1R}/\tau_{3R}$ in our simulations (Table 1) to the predictions of the Debye model[68,] which describes reorientation in molecular liquids as a random walk of a continuous angular variable (angular Brownian motion). The Debye model[68] for molecular reorientation estimates $\tau_{1R}/\tau_{2R}=3$ and $\tau_{1R}/\tau_{3R}=6$, at variance with our results for both High-$\zeta$ and Low-$\zeta$ molecules. The deviation persists even if we average High-$\zeta$ and Low-$\zeta$ correlation times with the average fractions of High-$\zeta$ and Low-$\zeta$ molecules, 0.36 and 0.64, respectively. We find the average values 4.8, 2.0, and 1.0 ps for $\tau_{1R}$, $\tau_{2R}$, and $\tau_{3R}$, respectively, in agreement with the overall average calculated by Guardia, Skarmoutsos, and Masia[44], which lead to $\tau_{1R}/\tau_{2R}=2.4$ and $\tau_{1R}/\tau_{3R}=4.8$. Hence, the Debye model is inadequate to describe the reorientation motions in liquid water under ambient conditions, in agreement with previous classical MD studies[69]. Here, we observe additionally that both High-$\zeta$ and Low-$\zeta$ molecules deviate from the Debye model, with the most significant deviations for the High-$\zeta$ molecules that are more tetrahedrally-ordered.

## 3.5 Translational-Vibrational Dynamics

For the translational-vibrational dynamics, for molecules that at $t=0$ are High-$\zeta$ or Low-$\zeta$, we calculate the TCFs

$$C_v^i(t) = \frac{\left\langle \vec{v}_i(0) \cdot \vec{v}_i(t) \right\rangle}{\left\langle \vec{v}_i(0)^2 \right\rangle} \tag{8}$$

of the velocity vectors $v_i$ of O and H atoms, where the index is $i$=O, H, and, via the Fourier transform, their associated spectral densities

$$S_v^i(\omega) = \int_0^\infty \cos(\omega \cdot t) \cdot C_v^i(t) \cdot dt . \tag{9}$$

We first consider the oxygen spectral density $S_v^O(\omega)$ (Figure 9). For both High-$\zeta$ and Low-$\zeta$ molecules, we find a low-frequency peak at 53 cm$^{-1}$ that in the literature is attributed to underlying mechanisms, e.g., hydrogen bridge bonds and cage effects[44,70,71]. In the range $\approx$ 125-200 cm$^{-1}$, we find differences between the High-$\zeta$ and Low-$\zeta$ $S_v^O(\omega)$. The spectrum is structured for High-$\zeta$ molecules—with a shoulder at 125 cm$^{-1}$, a minimum at 183 cm$^{-1}$, and a peek at 210 cm$^{-1}$—while it has only a broad shoulder or Low-$\zeta$. Another shoulder is observed for High-$\zeta$ molecules at higher frequencies around 275 cm$^{-1}$. The previous studies[44] attribute the bands between 125-180 cm$^{-1}$ to the intermolecular stretching-vibrational modes of hydrogen-bonded water dimers[44,71,72]. Furthermore, experiments[73] suggest that the intermolecular vibrations between 190 and 220 cm$^{-1}$ can be attributed to hindered translations of water molecules within a tetrahedral network. Moreover, a peak at 203 cm$^{-1}$ has been observed in experiments for supercooled water and attributed to the tetrahedral structure[74]. Hence, our results are consistent with a strong tetrahedral arrangement for High-$\zeta$ molecules and a more disordered structure for Low-$\zeta$ water.

We now consider the hydrogen spectral densities $S_v^H(\omega)$ (Figure 10). For both High-$\zeta$ and Low-$\zeta$ water molecules, we find the peak corresponding to the intramolecular H-O-H angle bending at 1587 cm$^{-1}$. However, at lower frequencies, in the librational region of the spectrum, we observe that High-$\zeta$ water molecules display two distinct peaks at 496 and 600 cm$^{-1}$, while Low-$\zeta$ only a broad peak with a maximum at 526 cm$^{-1}$. The peak at 600 cm$^{-1}$ is consistent with our findings revealing more hindered orientational dynamics for High-$\zeta$ molecules.

The $S_v^H(\omega)$ at high frequency (>3000 cm$^{-1}$), associated with intramolecular O-H stretching vibrations, is not symmetric and exhibits differences between the two types of molecules. It can be decomposed into three components—the central peak and two shoulders—each represented by a Voigt distribution function, i.e., the convolution of a Lorentzian and a Gaussian distribution[75].

We find that for High-$\zeta$ water molecules, the three components have peaks at 3192, 3329, and 3466 cm$^{-1}$, while for Low-$\zeta$ their maxima are at 3122, 3325, and 3486 cm$^{-1}$ (Figure 11). Hence, the lowest frequency peak of Low-$\zeta$ molecules is redshifted by 70 cm$^{-1}$. Moreover, the amplitude of the highest frequency distribution for Low-$\zeta$ water is smaller than for High-$\zeta$, resulting in a different shape.

Interestingly, the authors of previous infrared spectroscopic studies of liquid water[76] decomposed the O-H stretching band into three Gaussian curves. They assigned i) the lowest frequency Gaussian to molecules with an H-bond coordination number close to four; ii) the intermediate-frequencies Gaussian component to molecules with an average degree of connection larger than dimers or trimers but lower than those participating in percolating networks; iii) the highest-frequency Gaussian to water molecules being poorly connected to their environment.

Here, by integrating the areas of each Voigt distribution function, we find that for the High-$\zeta$ water molecules the low, intermediate, and high-frequency components are 58.5 %, 30.0 %, and 11.5 % of the total, while for Low-$\zeta$ they are 32.1 %, 64.2 %, and 3.7 %, respectively. Hence, most of the High-$\zeta$ molecules population has a coordination number of about four, while the Low-$\zeta$ population comprises less connected molecules.

### 3.6 Density Fluctuations from Molecular Collective Motions

To investigate the collective dynamic of the different environments, we calculate for both High-$\zeta$ and Low-$\zeta$ water molecules the *distinct* van Hove space-time correlation functions[77] $G_d^{OO}(r,t)/\rho$ where $\rho$ is the number density of water molecules and

$$G_d^{OO}(r,t) = \frac{1}{N}\left\langle \sum_i \sum_{j \neq i} \delta\left[r - |r_i(t) - r_j(0)|\right] \right\rangle \tag{10}$$

is proportional to the probability that a molecule is at position *r* at time *t*, given that a *distinct* molecule was at the origin (*r = 0*) at time *t = 0* and was High-$\zeta$ or Low-$\zeta$. This quantity measures the density fluctuations arising from collective molecular motions at different lengths- and timescales, and at *t=0* is $G_d^{OO}(r,0)/\rho = g^{OO}(r)$ (Figure 2).

We find that, for High-$\zeta$ water molecules, the first peak of the $G_d^{OO}(r,t)/\rho$ decreases in time, with a decrement of $\approx$54% within a very short time of 0.1 ps (Figure 12). On the other hand, the second maximum at $\approx$4.5 Å is stable up to the longest time investigated, 0.5 ps, becoming higher than the first for *t > 0.2* ps. Furthermore, the position of the first minimum shifts from $\approx$3.23 to 3.4 Å, but the first two peaks remain separated.

For Low-$\zeta$ molecules, the first peak decreases by $\approx$66% within 0.1 ps, while the second at r $\approx$ 4.5 Å is stable in time (Figure 11). However, in this case, the minimum at 3.48 Å is not shifted until it flattens out after t$\approx$0.5 ps, giving rise to a plateau extending up to the second maximum, consistent with the merging of the first two coordination shells.

Hence, the High-$\zeta$ water molecules' first and second coordination shells are preserved longer than those for the Low-$\zeta$ environment. We interpret this longer correlation time for the High-$\zeta$ structures as due to their cohesive tetrahedrality with a sharp separation at *t=0* between the first and second coordination shells.

Interestingly, the Low-$\zeta$ $G_d^{OO}(r,t)/\rho$ has isosbestic points at r $\approx$ 3.4 and 4.5 Å, and the High-$\zeta$ function has its first minimum converging at r $\approx$ 3.4 Å for long times. Therefore, it is reasonable to associate the two characteristic intermolecular distances 3.4 and 4.5 Å to the High-$\zeta$ and Low-$\zeta$ structures, consistent with the two-state models[56], or X-Ray[78] and neutron scattering[29] experiments for water.

## 4. Conclusions

We investigate the local structural order in ambient liquid water by the Car-Parrinello *ab initio* MD. The decomposition of the distributions of the local and coarse-grained order parameter $\zeta$ reveals the existence of two components or states exhibiting high and low coordination-shell separation. The two components, High-$\zeta$ and Low-$\zeta$ water molecules have different local environments. The High-$\zeta$ structure, with a sharp separation of the first and the second coordination shells, has a central water molecule forming four tetrahedral hydrogen bonds with its neighbors, with a $\approx$ 1:1 ratio of Low- to High-$\zeta$ bonded molecules. A Low-$\zeta$ molecule, on average, forms only three hydrogen bonds with its first coordination shell but has five neighbors, of which four are Low-$\zeta$, and one is High-$\zeta$, with a merging of the first and the second shell that induces structural disorder. Furthermore, the High-$\zeta$ molecules have a structure that correlates at least over two coordination shells, while the Low-$\zeta$ decorrelate at a much shorter distance. Therefore, High-$\zeta$ molecules have a higher probability of forming networks of tetrahedrally bonded molecules.

The use of *ab initio* MD allows us to calculate the dipole moment and reveal a relevant consequence of the difference in structure between the two water components. The High-$\zeta$ molecules have a dipole moment that is 6 % higher than the Low-$\zeta$ and almost equal to the molecular dipole moment of ice Ih. Therefore, our analysis reveals the presence of heterogeneities in the water dipole moment—hence, in the dielectric constant—that can extend further than two coordination shells at about 0.5 nm under ambient conditions. The time-dependent angular Van Hove self-correlation function calculation shows that the Low-$\zeta$ water molecules have more significant angular fluctuations than High-$\zeta$ at all the timescales considered here, up to 0.3 ps.

To estimate better over which timescales these heterogeneities are relevant, we calculate the first, the second, and the third-order Legendre time correlation functions (TCFs) for the orientation of each type of molecule. We find that the first, second and third order Legendre reorientational TCF i) decorrelate faster for Low-$\zeta$ water molecules and ii) exhibit larger deviations from the Debye model of molecular reorientation for High-$\zeta$ water molecules.

Furthermore, we analyze the low-frequency intermolecular oxygen vibrations of ambient High-ζ and the Low-ζ water, the librational low-frequency and the high-frequency part—associated with the intramolecular O-H bond-stretching—of the hydrogen spectral densities, $S_v^H(\omega)$. We find differences consistent with a tetrahedral network of hydrogen bonds coexisting with less connected Low-ζ water molecules.

We investigate the dynamic evolution of these different local structural networks in terms of the distinct van Hove space-time correlation function. We reveal that, while Low-ζ water loses its initial correlation within 0.2 ps, a longer time is required for High-ζ water molecules.

The present study also reveals the existence of two different categories of isosbestic points. The first category is related to the RDFs between the different types of water molecules. The crossing points are located at about 2.9, 3.8, and 5.15 Å, which coincide with the temperature-induced isosbestic points of the experimentally (X-Ray) measured RDF reported in the previous studies. The second category relates to the distinct part of the Van Hove space-time correlation functions, and the crossing points are located at about 3.4 and 4.5 Å. When correlated with previously reported two-state models[56] and experimental X-Ray[78] and neutron scattering studies of water[29], our results, by DFT-based CPMD simulations, provide strong evidence of the existence of two different types of local structures in ambient liquid water with implications for dipole moment fluctuations at the nm length scale and, at least, ps timescale, calling for new and definitive experiments confirming the two-state model.

**Acknowledgments**

We thankfully acknowledge the computer resources, technical expertise, and assistance provided by the Barcelona Supercomputing Center - Centro Nacional de Supercomputación. I.S. acknowledges the use of the computational facilities of the Computer Simulation in Condensed Matter Research Group (SIMCON) at the Physics Department of the Technical University of Catalonia (UPC) to analyze the CPMD simulation trajectories. G.F. and E.G. acknowledge support by Spanish grant PGC2018-099277-B-C22 and PGC2018-099277-B-C21 (MCIN/AEI/10.13039

/501100011033/ERDF "A way to make Europe"), respectively. G.F. acknowledges the support of ICREA Foundation (ICREA Academia prize).

**References**

[1] P. Gallo, K. Amann-Winkel, C.A. Angell, M.A. Anisimov, F. Caupin, C. Chakravarty, E. Lascaris, T. Loerting, A.Z. Panagiotopoulos, J. Russo, J.A. Sellberg, H.E. Stanley, H. Tanaka, C. Vega, L. Xu, L.G.M. Pettersson, Chem. Rev. 116 (2016) 7463.

[2] P. Gallo, J. Bachler, L.E. Bove, R. Böhmer, G. Camisasca, L.E. Coronas, H.R. Corti, I. de Almeida Ribeiro, M. de Koning, G. Franzese, V. Fuentes-Landete, C. Gainaru, T. Loerting, J.M. Montes de Oca, P.H. Poole, M. Rovere, F. Sciortino, C.M. Tonauer, G.A. Appignanesi, Eur. Phys. J. E 44 (2021) 143.

[3] C. A. Angell, in: F. Franks (Ed.), Water: A Comprehensive Treatise, Plenum, New York, 1982, Vol. 7.

[4] F. Franks, Water: A Matrix of Life, Royal Society of Chemistry, Cambridge, 2000.

[5] P.G. Debenedetti, J. Phys.: Condens. Matter 15 (2003) R1669.

[6] P.G. Debenedetti, H. E. Stanley, Phys. Today 56 (2003) 40.

[7] P.G. Debenedetti, Metastable Liquids: Concepts and Principles, Princeton University Press, Princeton,1996.

[8] A. Nilsson, L.G.M. Pettersson, Nat. Commun. 6 (2015) 8998.

[9] P. Ball, $H_2O$: A Biography of Water, Weidenfeld & Nicolson, London, 1999.

[10] G.W. Robinson, S.B. Zhu, S. Singh, M.W. Evans, Water in Biology, Chemistry, and Physics: Experimental Overviews and Computational Methodologies, World Scientific, Singapore, 1996.

[11] K. Stokely, M. G. Mazza, H. E. Stanley, and G. Franzese, Proc. Natl Acad. Sci. USA 107 (2010), 1301.

[12] B. Bagchi, Water in Biological and Chemical Processes: From Structure and Dynamics to Function, Cambridge University Press, Cambridge, 2013.

[13] V. Bianco, G. Franzese, Phys. Rev. Lett. 115 (2015) 108101.

[14] V. Bianco, G. Franzese, C. Dellago, I. Coluzza, Phys. Rev. X 7 (2017) 021047.

[15] F. Martelli, C. Calero, and G. Franzese, Biointerphases 16 (2021) 020801.

[16] H. R. Corti, G. A. Appignanesi, M. C. Barbosa, J. R. Bordin, C. Calero, G. Camisasca, M. D. Elola, G. Franzese, P. Gallo, A. Hassanali, K. Huang, D. Laria, C. A. Menendez, J. M. Montes de Oca, M. P. Longinotti, J. Rodriguez, M. Rovere, D. Scherlis, I. Szleifer, Eur. Phys. J. E 44 (2021) 136.

[17] F. Leoni, C. Calero, G. Franzese, ACS Nano 15 (2021) 19864.

[18] F. de los Santos, G. Franzese, Phys. Rev. E 85 (2012) 010602(R).

[19] M. Sasaki, Physica A 285 (2000) 315.

[20] C. Huang, K.T. Wikfeldt, T. Tokushima, D. Nordlund, Y. Harada, U. Bergmann, M. Niebuhr, T. M. Weiss, Y. Horikawa, M. Leetmaa, M. P. Ljungberg, O. Takahashi, A. Lenz, L. Ojamäe, A.P. Lyubartsev, S. Shin, L.G.M. Pettersson, A. Nilsson, Proc. Natl. Acad. Sci. U. S. A. 106 (2009) 15214.

[21] A. Nilsson, L.G.M. Pettersson, Chem. Phys. 389 (2011) 1.

[22] H.E. Stanley, S.V. Buldyrev, G. Franzese, N. Giovambattista, F.W. Starr, Phil. Trans. R. Soc. A 363 (2005) 509.

[23] G.N.I. Clark, G.L. Hura, J. Teixeira, A.K. Soper, T. Head-Gordon, Proc. Natl Acad. Sci. USA 107 (2010) 14003.

[24] G.N.I. Clark, C.D. Cappa, J.D. Smith, R.J. Saykally, T. Head-Gordon, Mol. Phys. 108 (2010) 1415.

[25] J. Niskanen, M. Fondell, C.J. Sahle, S. Eckert, R.M. Jay, K. Gilmore, A. Pietzsch, M. Dantz, X. Lu, D.E. McNally, T. Schmitt, V. Vaz da Cruz, V. Kimberg, F. Gel'mukhanov, A. Föhlisch, Proc. Natl. Acad. Sci. U.S.A. 116 (2019) 4058.

[26] L.G.M. Pettersson, Y. Harada, A. Nilsson, Proc. Natl. Acad. Sci. U.S.A. 116 (2019) 17156.

[27] J. Niskanen, M. Fondell, C.J. Sahle, S. Eckert, R.M. Jay, K. Gilmore, A. Pietzsch, M. Dantz, X. Lu, D.E. McNally, T. Schmitt, V. Vaz da Cruz, V. Kimberg, F. Gel'mukhanov, A. Föhlisch, Proc. Natl. Acad. Sci. U.S.A. 116 (2019) 17158.

[28] E. Duboué-Dijon, D. Laage, J. Phys. Chem. B 119 (2015) 8406.

[29] A.K. Soper, M.A. Ricci, Phys. Rev. Lett. 84 (2000) 2881.

[30] Z. Yan, S.V. Buldyrev, P. Kumar, N. Giovambattista, P.G. Debenedetti, H.E. Stanley, Phys. Rev. E 76 (2007) 051201.

[31] E. Shiratani, M. Sasai, J. Chem. Phys. 108 (1998) 3264.

[32] J.R. Errington, P.G. Debenedetti, Nature 409 (2001) 318.

[33] R.H. Henchman, S.J. Cockram, Faraday Discuss. 167 (2013) 529.

[34] T.A. Kesselring, E. Lascaris, G. Franzese, S.V. Buldyrev, H.J. Herrmann, H.E. Stanley, J. Chem. Phys. 138 (2013) 244506.

[35] I. Skarmoutsos, M. Masia, E. Guardia, Chem. Phys. Lett. 648 (2016) 102.

[36] J. Russo, H. Tanaka, Nat. Commun. 5 (2014) 3556.

[37] R. Shi, H. Tanaka, J. Chem. Phys. 148 (2018) 124503.

[38] F. Martelli, J. Chem. Phys. 150 (2019) 094506.

[39] F. Martelli, J. Crain, G. Franzese, ACS Nano 14 (2020) 8616.

[40] H. Tanaka, H. Tong, R. Shi, J. Russo, Nat. Rev. Phys. 1 (2019) 333.

[41] R. Shi, J. Russo, H. Tanaka, J. Chem. Phys. 149 (2018) 224502.

[42] R. Shi, J. Russo, H. Tanaka, Proc. Natl. Acad. Sci. U. S. A. 115 (2018) 9444.

[43] R. Shi, H. Tanaka, J. Am. Chem. Soc. 142 (2020) 2868.

[44] E. Guardia, I. Skarmoutsos, M. Masia, J. Phys. Chem. B 119 (2015) 8926.

[45] J. Yang, R. Dettori, J.P.F. Nunes, N.H. List, E. Biasin, M. Centurion, Z. Chen, A.A. Cordones, D.P. Deponte, T.F. Heinz, M.E. Kozina, K. Ledbetter, M.-F. Lin, A.M.

Lindenberg, M. Mo, A. Nilsson, X. Shen, T.J.A. Wolf, D. Donadio, K.J. Gaffney, T.J. Martinez, X. Wang, Nature 596 (2021) 531.

[46] A. Becke, Phys. Rev. A 38 (1988) 3098.

[47] C. Lee, W. Yang, R. Parr, Phys. Rev. B 37 (1988) 785.

[48] O.A. von Lilienfeld, I. Tavernelli, U. Röthlisberger, D. Sebastiani, Phys. Rev. Lett. 93 (2004) 153004.

[49] I.-C. Lin, A.P. Seitsonen, M.D. Coutinho-Neto, I. Tavernelli, U. Röthlisberger, J. Phys. Chem. B 113 (2009) 1127.

[50] N. Troullier, J.L. Martins, Phys. Rev. B 43 (1991) 1993.

[51] O. Akin-Ojo, Y. Song, F. Wang, J. Chem. Phys. 129 (2008) 064108.

[52] I.-F.W. Kuo, C.J. Mundy, M.J. McGrath, J.I. Siepmann, J. VandeVondele, M. Sprik, J. Hutter, B. Chen, M.L. Klein, F. Mohamed, M. Krack, M. Parrinello, J. Phys. Chem. B 108 (2004) 12990.

[53] I. Skarmoutsos, A. Henao, E. Guardia, J. Samios, J. Phys. Chem. B 125 (2021) 10260.

[54] A. Soper, Chem. Phys. 258 (2000) 121.

[55] I. Skarmoutsos, E. Guardia, J. Samios, J. Chem. Phys. 133 (2010) 014504.

[56] G.W. Robinson, C.H. Cho, J. Urquidi, J. Chem. Phys. 111 (1999) 698.

[57] M. Greger, M. Kollar, D. Vollhardt, Phys. Rev. B 87 (2013) 195140.

[58] N. Marzari, D. Vanderbilt, Phys. Rev. B 56 (1997) 12847.

[59] P.L. Silvestrelli, N. Marzari, D. Vandebilt, M. Parrinello, Solid State Commun. 107 (1998) 7.

[60] P.L. Silvestrelli, M. Parrinello, Phys. Rev. Lett. 82 (1999) 3308.

[61] D. Pan, L. Spanu, B. Harrison, D.A. Sverjensky, G. Galli, Proc. Natl. Acad. Sci. U. S.A. 110 (2013) 6646.

[62] E. Guardia, I. Skarmoutsos, M. Masia, J. Chem. Theory Comput. 5 (2009) 1449.

[63] Y.S. Badyal, M.-L. Saboungi, D.L. Price, S.D. Shastri, D.R. Haeffner, A.K. Soper, J. Chem. Phys. 112 (2000) 9206.

[64] E.R. Batista, S.S. Xantheas, H. Jónsson, J. Chem. Phys. 109 (1998) 4546.

[65] I. Bakó, J. Daru, S. Pothoczki, L. Pusztai, K. Hermansson, J. Mol. Liq. 293 (2019) 11579.

[66] I. Bakó, I. Mayer, J. Phys. Chem. A 120 (2016) 4408.

[67] I. Skarmoutsos, S. Mossa, E. Guardia, J. Chem. Phys. 150 (2019) 124506.

[68] P. Debye, Polar Molecules, Dover, New York, 1945.

[69] D. Laage, D., J.T. Hynes, J. Phys. Chem. B 112 (2008), 14230.

[70] M. Galvin, D. Zerulla, ChemPhysChem 12 (2011) 913.

[71] K.H. Tsai, T.M. Wu, Chem. Phys. Lett. 417 (2006) 389.

[72] B.S. Mallik, A. Semparithi, A. Chandra, J. Phys. Chem. A 112 (2008) 5104.

[73] R. Schwan, C. Qu, D. Mani, N. Pal, G. Schwaab, J.M. Bowman, G.S. Tschumper, M. Havenith, Angew. Chem. Int. Ed. 59 (2020) 11399.

[74] S. Funke, F. Sebastiani, G. Schwaab, M. Havenith, J. Chem. Phys. 150 (2019) 224505.

[75] P.M.A. Sherwood, Surf. Interface Anal. 51 (2019) 254.

[76] J.-B. Brubach, A. Mermet, A. Filabozzi, A. Gerschel, P. Roy, P. J. Chem. Phys. 122 (2005) 184509.

[77] J.-P. Hansen, I.R. McDonald, Theory of Simple Liquids, 3rd ed., Academic Press, London, 2006.

[78] L. Fu, A. Bienenstock, S. Brenan, J. Chem. Phys. 131 (2009) 234702.

## TABLES

**Table 1:** The first, second, and third-order Legendre orientational correlation times and their ratios for water molecules that at time *t=0* are in the High-$\zeta$ and Low-$\zeta$ states, as calculated from Eq.(7).

| Type | $\tau_{1R}$ (ps) | $\tau_{2R}$ (ps) | $\tau_{3R}$ (ps) | $\tau_{1R}/\tau_{2R}$ | $\tau_{1R}/\tau_{3R}$ |
|---|---|---|---|---|---|
| High-$\zeta$ | 5.3 | 2.3 | 1.3 | 2.3 | 4.1 |
| Low-$\zeta$ | 4.5 | 1.8 | 0.9 | 2.5 | 5.0 |

## FIGURES

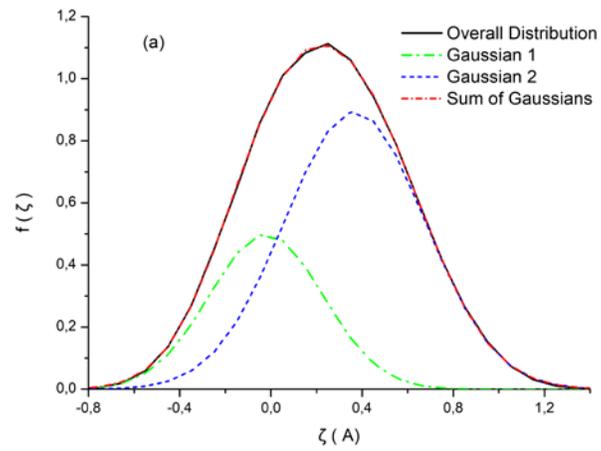

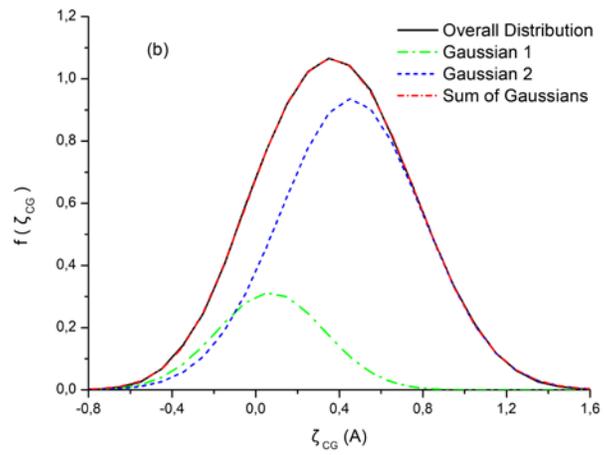

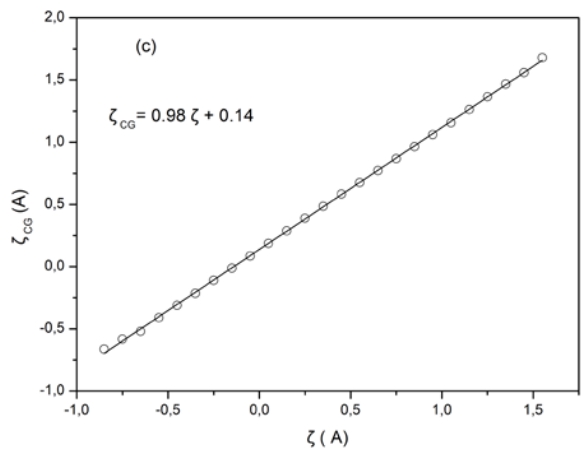

**Figure 1**

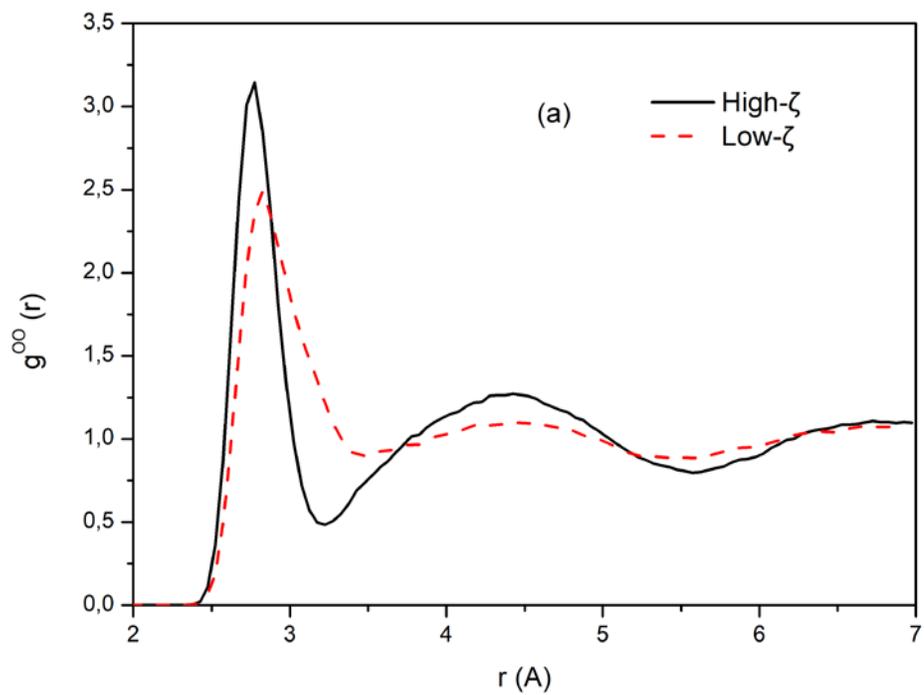

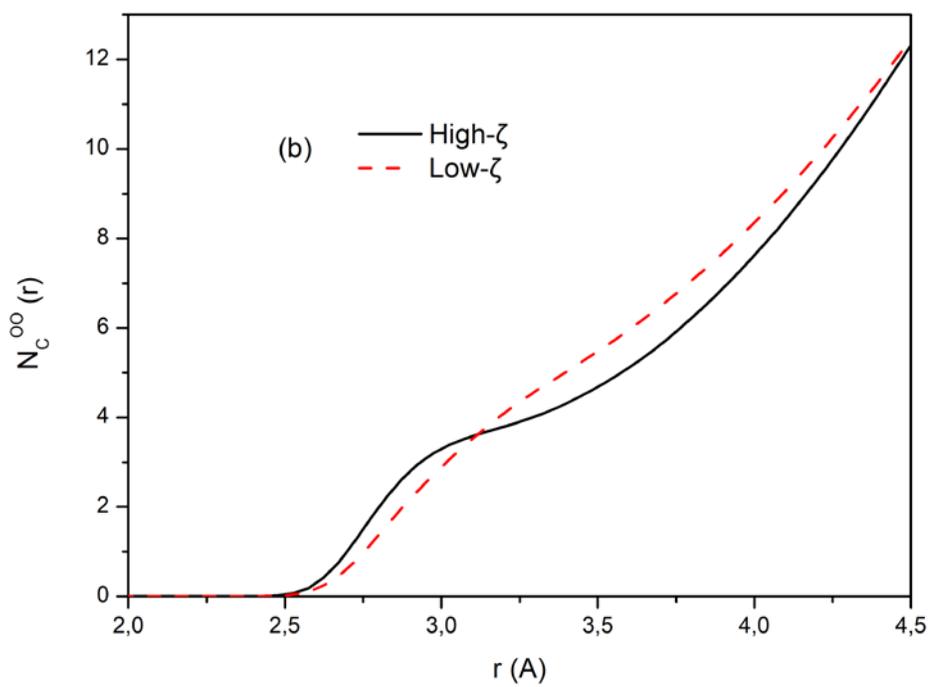

**Figure 2**

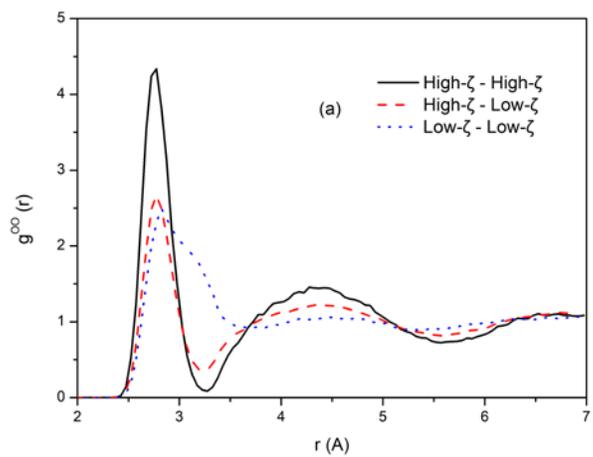

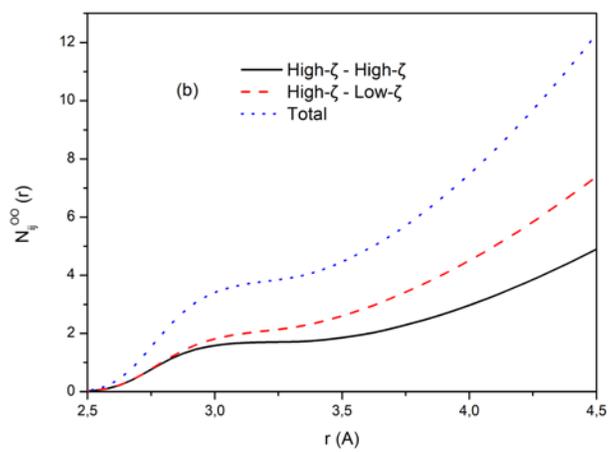

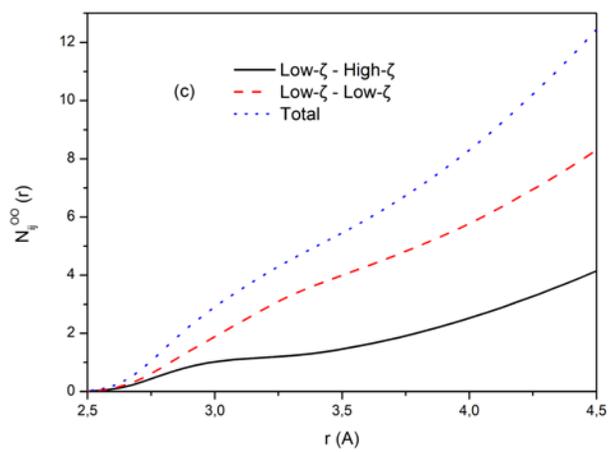

**Figure 3**

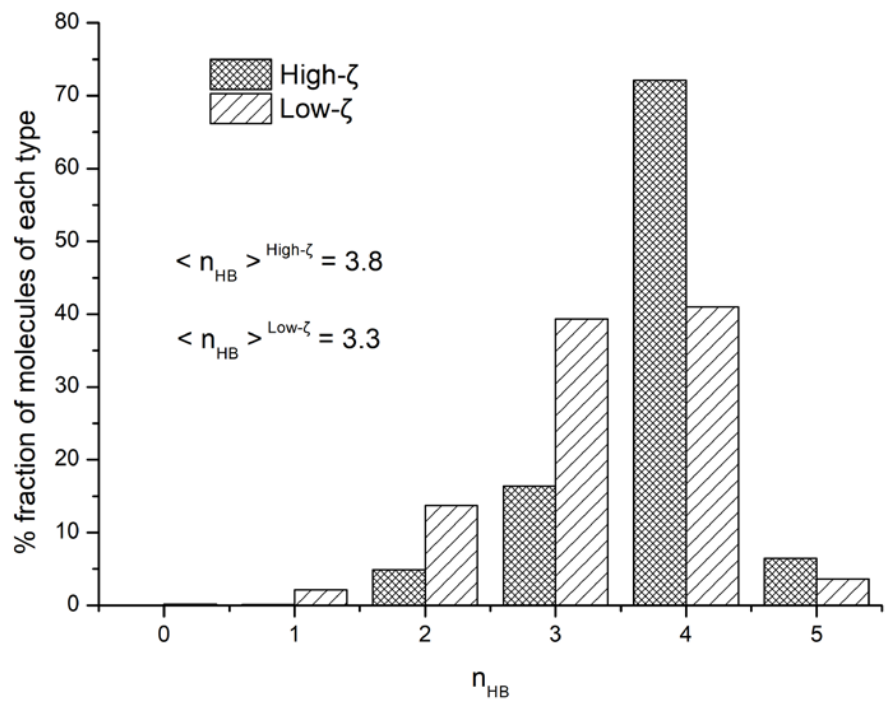

**Figure 4**

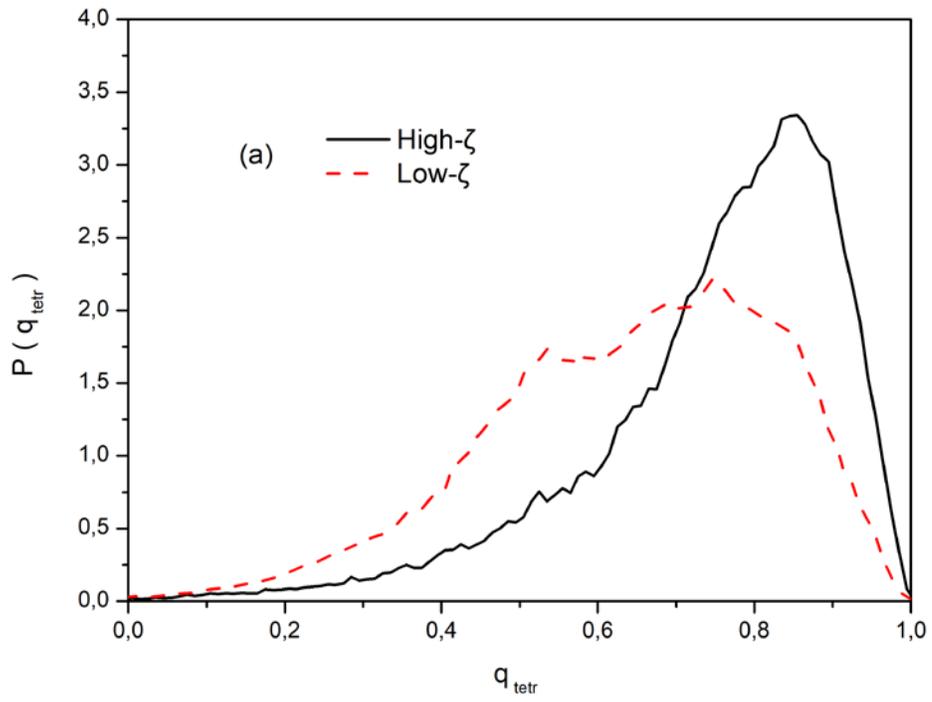

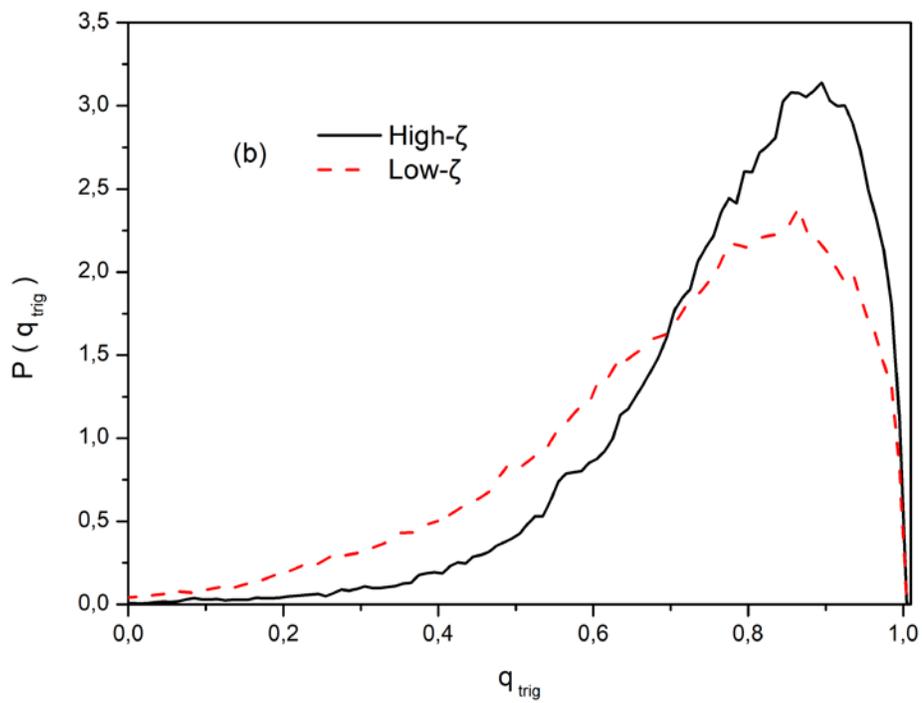

**Figure 5**

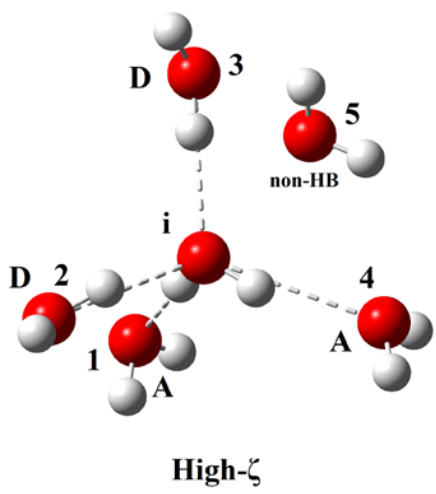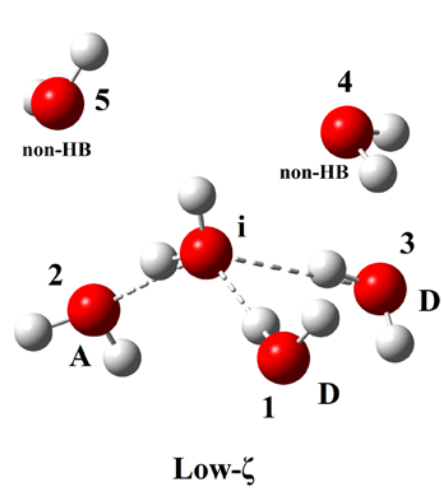

**Figure 6**

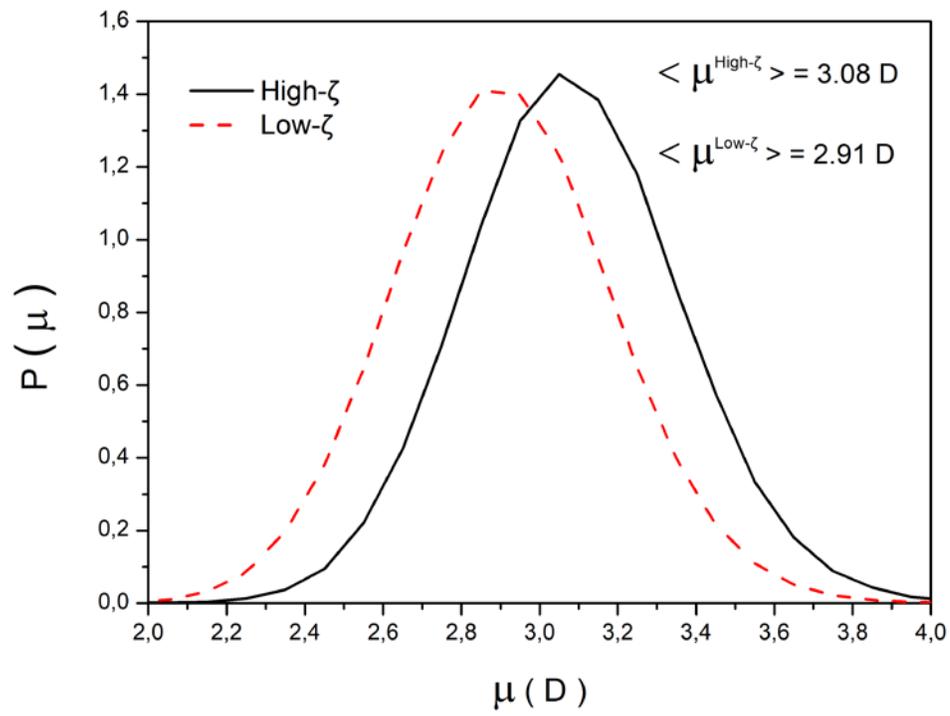

**Figure 7**

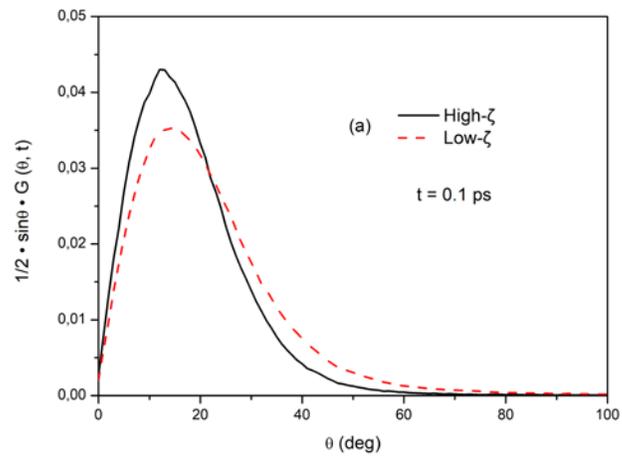

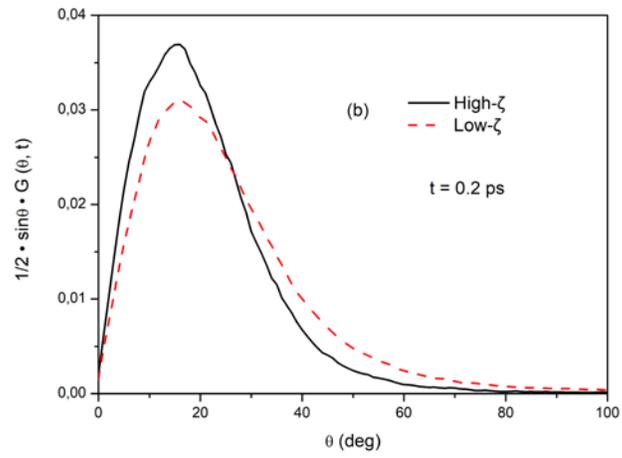

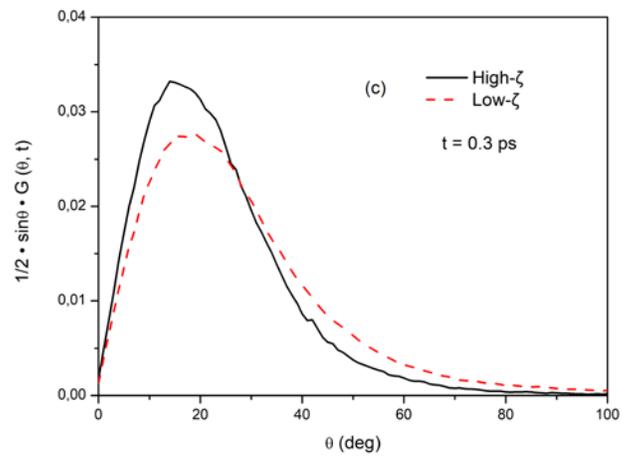

**Figure 8**

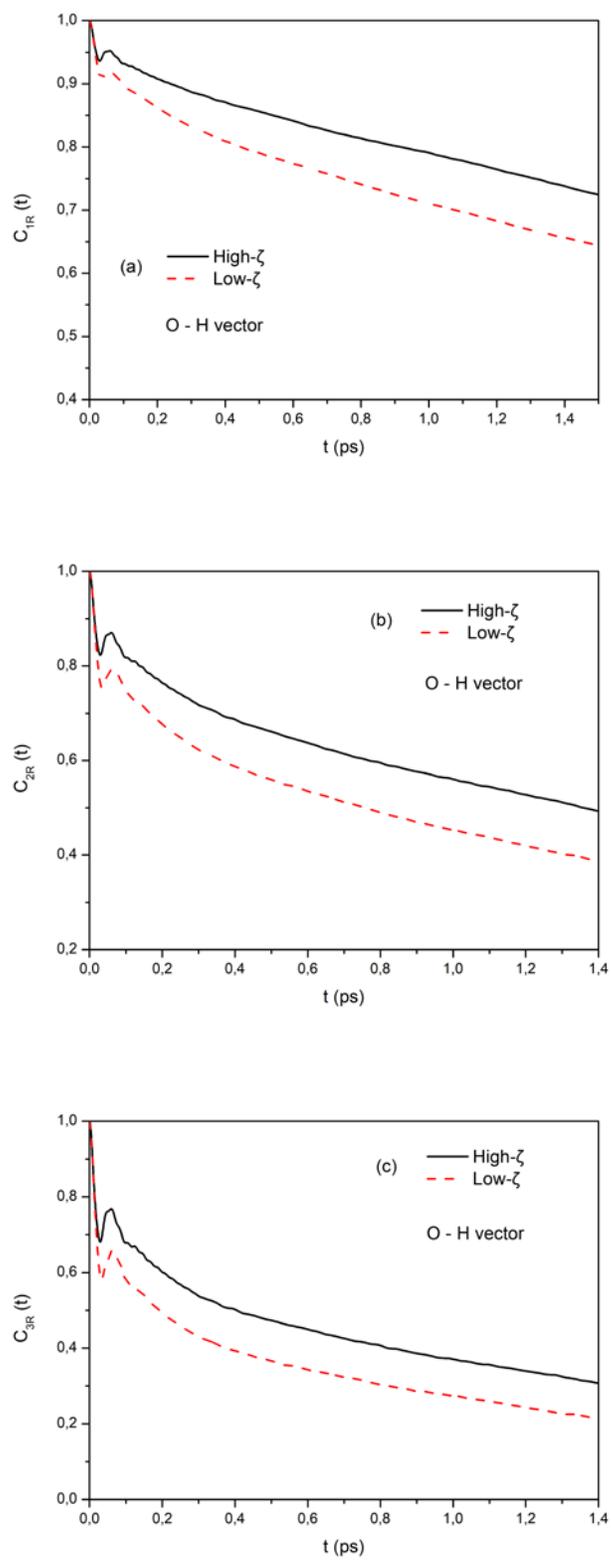

**Figure 9**

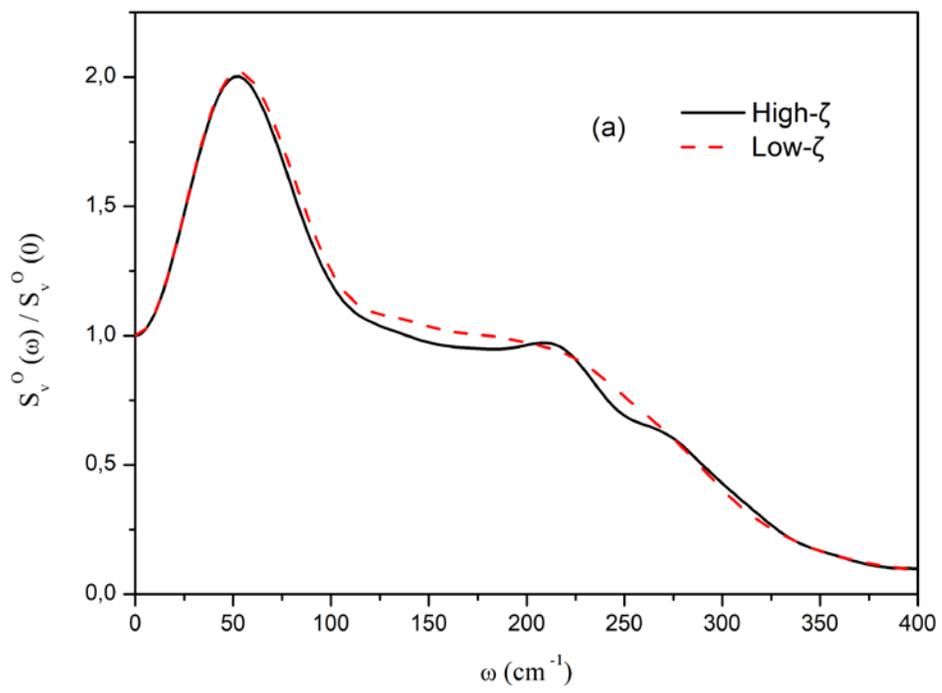

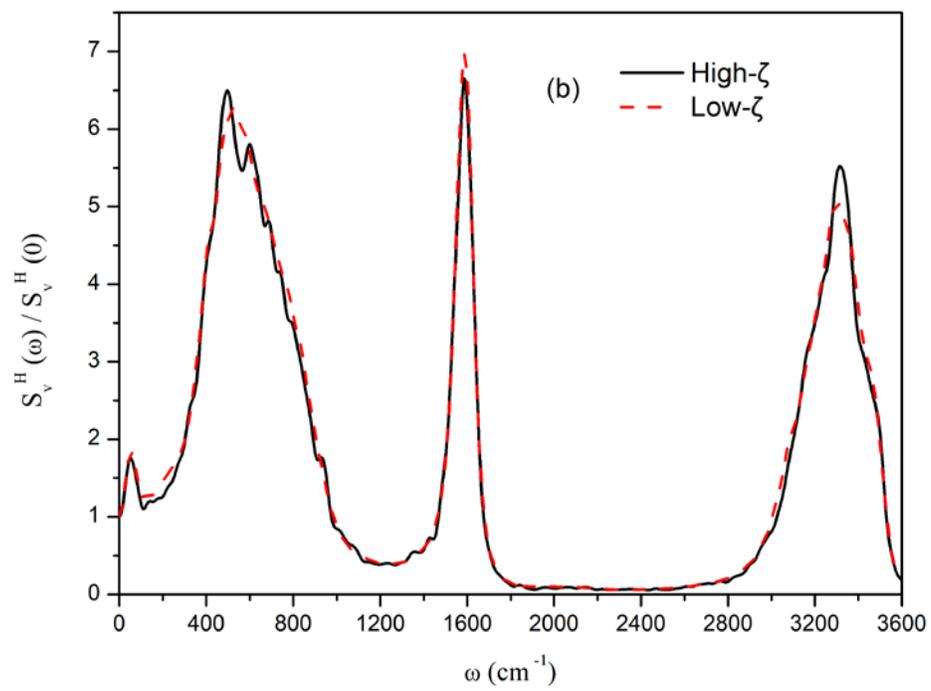

**Figure 10**

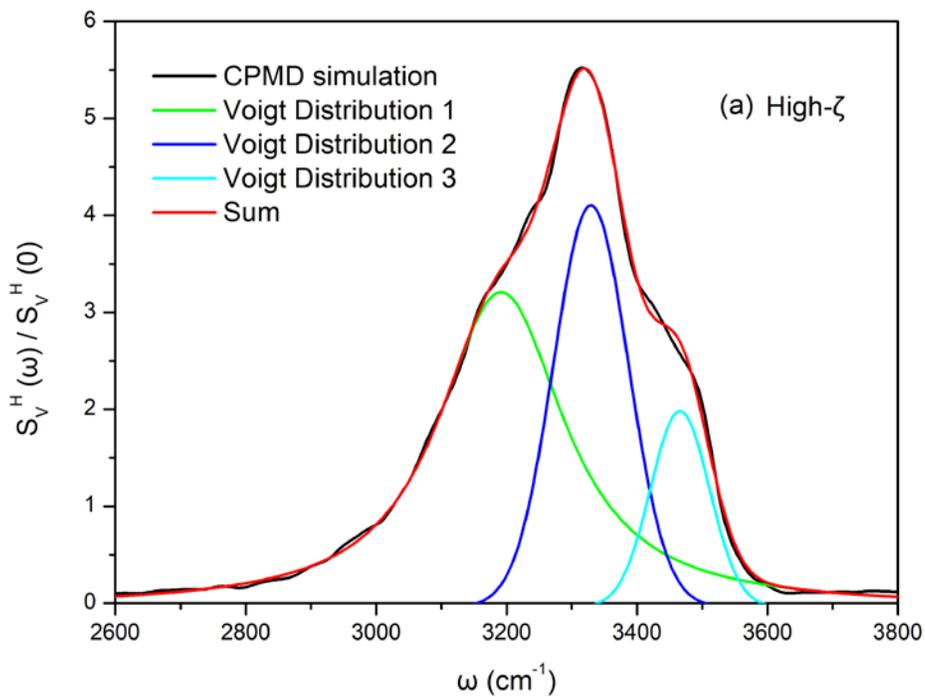

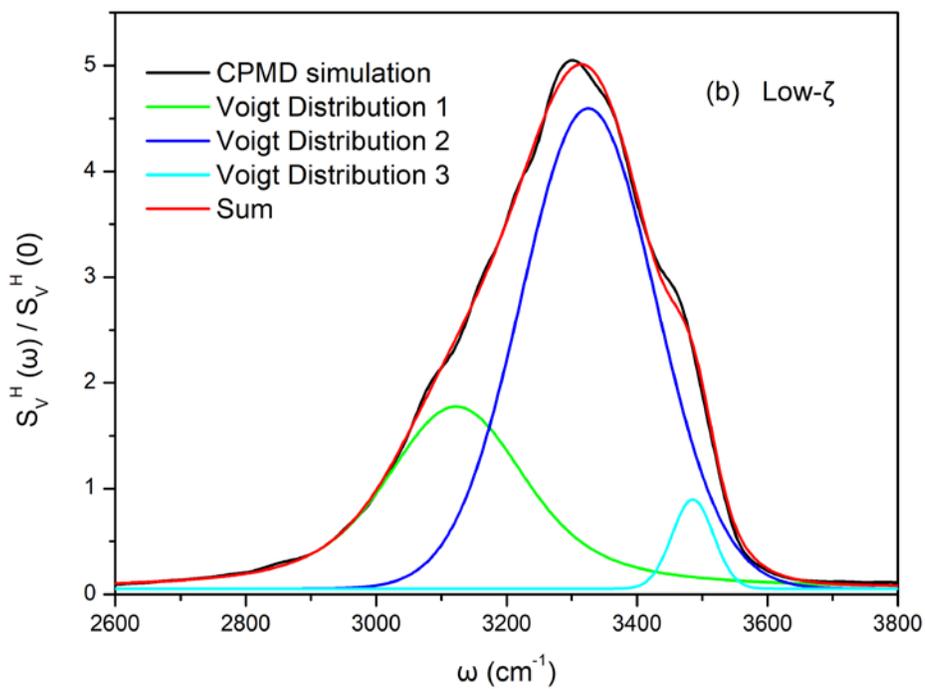

**Figure 11**

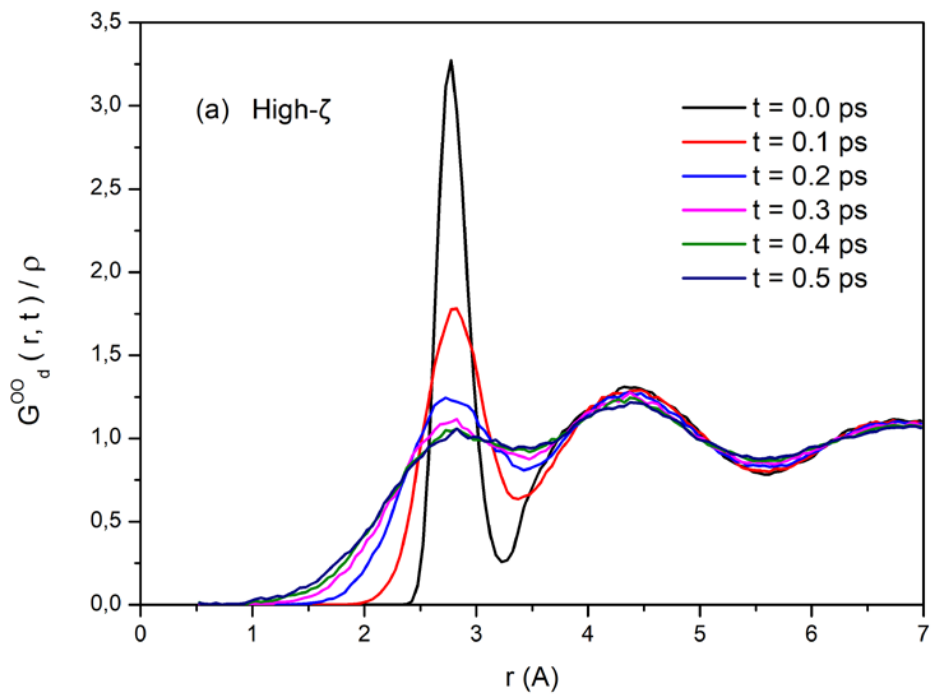

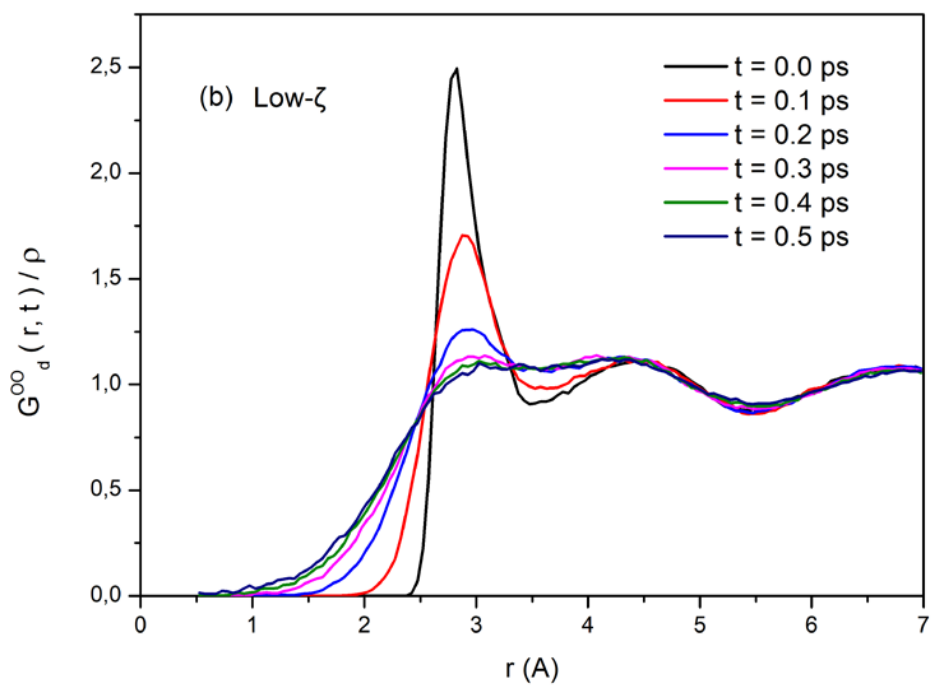

**Figure 12**

**FIGURE CAPTIONS**

**Figure 1: Water under ambient conditions has two structural components.** The normalized probability density distributions (solid black line) for (a) the local structural descriptor $\zeta$, and (b) its coarse-grained average $\zeta_{CG}$ of water under ambient conditions, as calculated by our DFT-based MD simulations, can be decomposed into two Gaussian components (1 long-dash-dotted green and 2 dashed blue lines), whose sum (dash-dotted red line) is indistinguishable from the original data. (c) The parametric plot of $\zeta_{CG}$ vs. $\zeta$ (circles) shows that they are linearly related as $\zeta_{CG} = 0.98\,\zeta + 0.14$ (solid line).

**Figure 2: The two components, High-$\zeta$ and Low-$\zeta$ water molecules have different local environments.** (a) The O-O RDF, $g^{OO}$, as a function of the radial distance $r$, for High-$\zeta$ molecules (solid line) has two well-separated maxima, while for Low-$\zeta$ (dashed line), the intermediate minimum, marking the separation between the first and the second coordination shell, is less pronounced. (b) The coordination number $N_C^{OO}$ is higher for the Low-$\zeta$ (dashed line) than the High-$\zeta$ molecules (solid line) starting from the first coordination shell, consistent with a denser local environment.

**Figure 3: High-$\zeta$ water molecules have a sharp tetrahedral coordination shell with a ≈ 1:1 ratio of Low- to High-$\zeta$ neighbors under ambient conditions, while Low-$\zeta$ molecules have five neighbors, with a 4:1 ratio, merging the first and the second shell.** (a) The O-O RDF, $g^{OO}_{AB}$, as a function of the radial distance $r$, for $A$, $B$=High-$\zeta$ molecules (solid black line), $A$=High-$\zeta$ and $B$=Low-$\zeta$ (long-dashed red line), and $A$, $B$=Low-$\zeta$ (dashed blue line). (b) The coordination number $N^{OO}_{AB}$ for $A$=High-$\zeta$ molecule and $B$=High-$\zeta$ (solid black line) or $B$=Low-$\zeta$ (long-dashed red line), and their sum (dashed blue line). (c) As in (b) but for $A$=Low-$\zeta$ molecule.

**Figure 4: High-$\zeta$ molecules form about four hydrogen bonds, while Low-$\zeta$ only three.** Histogram of hydrogen bonds for High-$\zeta$ (darkly shaded bars) and Low-$\zeta$ (lightly shaded bars) water molecules under ambient conditions.

**Figure 5: High-$\zeta$ molecules have tetrahedral order, while Low-$\zeta$ do not.** The normalized probability distribution for (a) the tetrahedral order parameters for High-$\zeta$ water

molecules (continuous black profile) and Low-$\zeta$ (dashed red profile) are pretty different, while for (b) the trigonal order parameters are similar.

**Figure 6: Representative snapshots of two water molecules in the High-$\zeta$ and Low-$\zeta$ states**. In both panels, we show the central water molecule *i* with its first five neighbors, with indices 1-5 from the nearest to the farthest, and labels *A, D, non-HB* for acceptors, donors, and non-bonded, respectively. The coordination shell has a radius $r_{shell}$ and $r_{ij}$ is the distance between molecules *i* and *j*. (a) In the High-$\zeta$ state, the water molecule *i* has a coordination number $Nc^{OO}=4$, $n_{HB}=4$ hydrogen bonds, $q_{tetr} \approx 0.89$, and $\zeta \approx 0.6$ Å. (b) In the Low-$\zeta$ state, it has $Nc^{OO}=5$, $n_{HB}=3$, $q_{tetr} \approx 0.49$, and $\zeta \approx -0.04$ Å.

**Figure 7: High-$\zeta$ dipole moment is as in tetrahedral structures and higher than the Low-$\zeta$.** The normalized probability density distribution for the molecular dipole moment of High-$\zeta$ water molecules (continuous black profile) has a 6.5% higher average than the Low-$\zeta$ (dashed red profile).

**Figure 8: Low-$\zeta$ water molecules have more significant angular fluctuations than High-$\zeta$.** The average $1/2 \cdot \sin\theta \cdot G(\theta,t)$ of the time-dependent angular Van Hove self-correlation functions for Low-$\zeta$ molecules (dashed red profile), calculated at *t* = 0.1, 0.2, and 0.3 ps (panels a, b, c, respectively), is more populated at a large angle $\theta$ than for the High-$\zeta$ case (continuous black profile).

**Figure 9: The reorientational decorrelation of the Low-$\zeta$ water is 40% faster in time than High-$\zeta$ water under ambient conditions**. The average (a) first, (b) second, and (c) third order Legendre TCFs for the orientation of Low-$\zeta$ molecules (dashed red function) at ≈ 0.9 ps have the values reached by the High-$\zeta$ molecules (continuous black profile) at 1.5 ps.

**Figure 10: The spectral densities of atomic translation velocities are consistent with tetrahedral structures for High-$\zeta$ water and disordered arrangments for Low-$\zeta$.** (a) The normalized oxygen $S_v^O(\omega)/S_v^O(0)$ spectral densities for the atomic velocity-time correlation functions of High-$\zeta$ (continuous black profile) and Low-$\zeta$ water molecules (dashed red profile) differ in the range of frequencies (125≤ $\omega$ ≤220 cm$^{-1}$) associated with

tetrahedral structures. (b) The corresponding hydrogen $S_v^H(\omega)/S_v^H(0)$ spectral densities show a difference in the low-frequency part associated with the libration motion.

**Figure 11: The decomposition of the high-frequency spectral densities of H translation velocities shows that High-$\zeta$ water molecules have mainly four H-bonds and are more connected than Low-$\zeta$.** Our spectral density $S_v^H(\omega)/S_v^H(0)$ for (a) High-$\zeta$ and (b) Low-$\zeta$ water molecules, generated by CPMD simulations under ambient conditions (solid black curves), are decomposed into three Voigt distribution functions (1, 2, and 3, green, blue, and turquoise, respectively) whose sum (red) is indistinguishable from the simulation results. The majority component is the first for High-$\zeta$ and the second for Low-$\zeta$.

**Figure 12: The High-$\zeta$ water molecules keep their first and second coordination shells separated up to at least 0.5 ps while Low-$\zeta$ molecules merge them within 0.2 ps under ambient conditions.** The *distinct* van Hove space-time correlation function, $G_d^{OO}(r,t)/\rho$, calculated at $t = 0.0\text{-}0.5$ ps, for water molecules that at $t=0$ are (a) High-$\zeta$ and (b) Low-$\zeta$.